\documentclass[aps,amsmath,prl,twocolumn,superscriptaddress,nofootinbib]{revtex4-1}
\usepackage{graphicx}
\usepackage{bm}
\usepackage{epsfig}
\usepackage{amsmath,latexsym}
\usepackage{epstopdf}
\usepackage{bbold}
\newcommand{\nn}{\nonumber}

\newcommand{\beq}{\begin{equation}}
\newcommand{\eeq}{\end{equation}}
\newcommand{\bea}{\begin{eqnarray}}
\newcommand{\eea}{\end{eqnarray}}
\newcommand{\mo}{\mathcal{O}}
\newcommand{\vacproj}{|0\rangle \langle 0 |}

\begin{document}

\title{On the Annihilation Rate of WIMPs }

\author{Matthew Baumgart, Ira Z. Rothstein, and Varun Vaidya\vspace{0.4cm}}
\affiliation{Department of Physics, Carnegie Mellon University,
    Pittsburgh, PA 15213}


\begin{abstract}
We develop a formalism that allows one to systematically 
calculate the WIMP annihilation rate into gamma rays whose energy far exceeds
the weak scale. A factorization theorem is presented which 
separates the radiative corrections stemming from initial-state potential
interactions from loops involving the final state. This separation 
allows us to go beyond the fixed order calculation, which is polluted by
large infrared logarithms. For the case of Majorana WIMPs transforming in
the adjoint representation of $SU(2)$, we present the result for the resummed rate at leading
double-log accuracy in terms of two initial-state partial-wave matrix elements and one hard matching coefficient. For a given model, one may calculate the cross section by finding the tree level matching coefficient and determining the value of a local four-fermion
operator.  The effects of resummation can be as large as 100\% for a 20 TeV WIMP.  However, for lighter WIMP masses relevant for the thermal relic scenario, leading-log resummation modifies the Sudakov factors only at the 10\% level.  Furthermore, given comparably-sized Sommerfeld factors, the total effect of radiative corrections on the semi-inclusive photon annihilation rate is found to be percent-level.  The generalization of the formalism to other types of WIMPs is discussed.
\end{abstract}

\maketitle

The gravitational evidence for dark matter (DM) is overwhelming, but despite considerable effort we are still awaiting its definitive non-gravitational observation.
Searches continue both via direct as well as indirect methods.  In principle a convincing complete dark matter model could be built by combining
missing energy event rates at the LHC with underground detection and/or astronomical cosmic/gamma ray signatures.
The negative results of these searches to date has allowed us to eliminate large swaths of parameter space.
In particular,  the WIMP scenario within the Minimal Supersymmetric Standard Model (MSSM) has been highly constrained. At present, we have yet to see evidence of new weak-scale physics, which implies that if the MSSM, or some variant thereof, is correct, then the supersymmetry (SUSY) breaking scale may be considered as being uncomfortably tuned.  

Nonetheless, if dark matter has a straightforward connection to known physics, the thermal-relic, weakly interacting massive particle (WIMP) provided by SUSY is one of the most elegant candidates.  The possibility of obtaining such a particle from SUSY continues to be a principle argument for that scenario (along with gauge coupling unification).  Pure-wino dark matter is the most attractive candidate of  ``mini-split'' SUSY, which foregoes strict naturalness for simpler model-building \cite{minisplit}.  In a typical implementation, gravity-mediation generates sfermion masses at $\mo(100 \textendash 1000)$ TeV and anomaly-mediation \cite{amsb} gives gaugino masses of $\mo(1 \textendash 10)$ TeV, with a wino LSP.  

For a thermal relic, the measured dark matter abundance rules out bino DM and very narrowly constrains $M_{\rm Wino} \equiv (M_{\chi})$ = 2.7-2.9 TeV \cite{Fan:2013faa,constraint}.    A question of paramount importance then is whether the present annihilation rate to photons is sufficiently large  to rule out thermal wino dark matter by the non-observation of TeV-scale photon lines at the air Cherenkov telescope, HESS.  Some groups have recently claimed this to be the case unless our galaxy's DM profile is highly cored or if higher-loop corrections result in a significant $\sim \mo(1/{\rm few})$ decrease in the annihilation rate to $\gamma + X$ \cite{Cohen:2013ama,Fan:2013faa}.  It is the proper treatment of the latter that is our primary motivation.  While a TeV-scale wino provides motivation for a more precise calculation, we stress that our formalism is generic for any WIMP that produces energetic, observable particles.

As is well known, the WIMP annihilation rate receives large radiative corrections
beyond tree level, as the rate is sensitive to the infrared (IR) scale $M_W\sim 100$ GeV.
The existence of this scale in the rate leads the large corrections to take two forms, namely
$\alpha \frac{M_\chi^2}{M_W^2}$ and $\alpha \log^2(\frac{M_\chi^2}{M_W^2})$ $(\alpha \equiv \frac{g_2^2}{4\pi})$.
The former are due to potential interactions between the incoming
nonrelativistic particles, while the latter are a result of soft-collinear gauge
boson emission which do not cancel due to the non-singlet nature of the initial states \cite{Ciafoloni}. The resummation of the potential exchange is handled by solving
the appropriate Schr\"{o}dinger equation. However, the resummation of the
large logs is not accomplished with such ease. Furthermore, once one considers
both such interactions simultaneously, one must understand how these two
pieces of physics decouple, if at all. Thus, the uncertainties in the tree level
annihilation rate can be considered to be of order one-hundred percent until
these singular radiative corrections are tamed in a systematic fashion.

Given that the large logs are a consequence
of dealing with multiple scales simultaneously, 
resummations are typically accomplished by first factorizing the rates.  
Such a factorization will also allow for the explicit separation of the
potential corrections from the IR radiation.
An efficient way of approaching  factorization is through the use
of effective field theory (EFT). For this particular application we need to
construct an EFT which can treat a process in which there are static massive particles (WIMPs) annihilating 
into highly-energetic particles. We will concentrate on photons as final states, though one can generalize our work
here to the case of fermions as well. To treat these various kinematic regimes we utilize a hybrid effective
theory which combines ideas from Non-Relativistic QCD (NRQCD) \cite{NRQCD}, which is used to
study quarkonia, with Soft-Collinear Effective Theory (SCET) \cite{SCET}, which can be utilized
to prove factorization theorems in high energy scattering \cite{SCETHE}. Similar, though distinct,  hybrid theories
have been utilized to study the photon spectrum in radiative onium decays \cite{Leibovich}.  SCET in the context of electro-weak theory has been explored in \cite{aneesh}.  Here we will not go into the details of the effective theory, which will be left for a separate publication \cite{future}.

We begin by enumerating the relevant modes which we will factorize. The dark matter are assumed to have velocities on
the order of $10^{-3}$, and as such will be static for our purposes. In principle, finite velocity corrections
could be included, but until an actual detection is made such corrections are not a pressing matter.
Furthermore, we will take the detected photon  ($E_\gamma$) to have energies far greater then the weak scale,
and ignore corrections which scale as powers  of $M_W/E_\gamma$. 
Thus, our power counting parameter will be $\lambda\equiv M_W/ (E_\gamma \sim M_\chi)$.

Given these kinematics there
are four relevant modes in the problem. The WIMP fields, whose energy and momenta scale as $\lambda^2$ and $\lambda$
respectively (all dimensions are in units of $M_\chi$), collinear fields whose light-cone coordinate momentum scales as
($k_+\sim 1,k_-\sim \lambda^2, k_\perp \sim \lambda)$, soft gauge boson fields with  ($k_+\sim \lambda,k_-\sim \lambda, k_\perp \sim \lambda)$ 
and potential gauge boson and Higgs fields whose energy and momenta scale as $\lambda^2$ and $\lambda$ respectively.
The potential modes will not play a role in the resummation. For the processes of interest here, the Higgs boson
will not play a role at leading order in $\lambda$.  Its effects will be discussed in \cite{future}.

At leading order in $\lambda$ we may work in the unbroken phase of the theory with impunity.
We will thus treat all the gauge bosons as massless, save for the times when the mass will
be needed to cutoff the IR singularities. Furthermore, for our purposes we may ignore the 
$U(1)$ factor of the SM gauge group as it will not play a role until higher orders. 
Thus, after integrating out the hard modes with invariant mass scales of order $M_\chi$, we
are left with an effective theory composed of nonrelativistic $\chi$ fields, and collinear and
soft $SU(2)$ gauge bosons.

\begin{figure}
\centerline{\scalebox{0.50}{\includegraphics{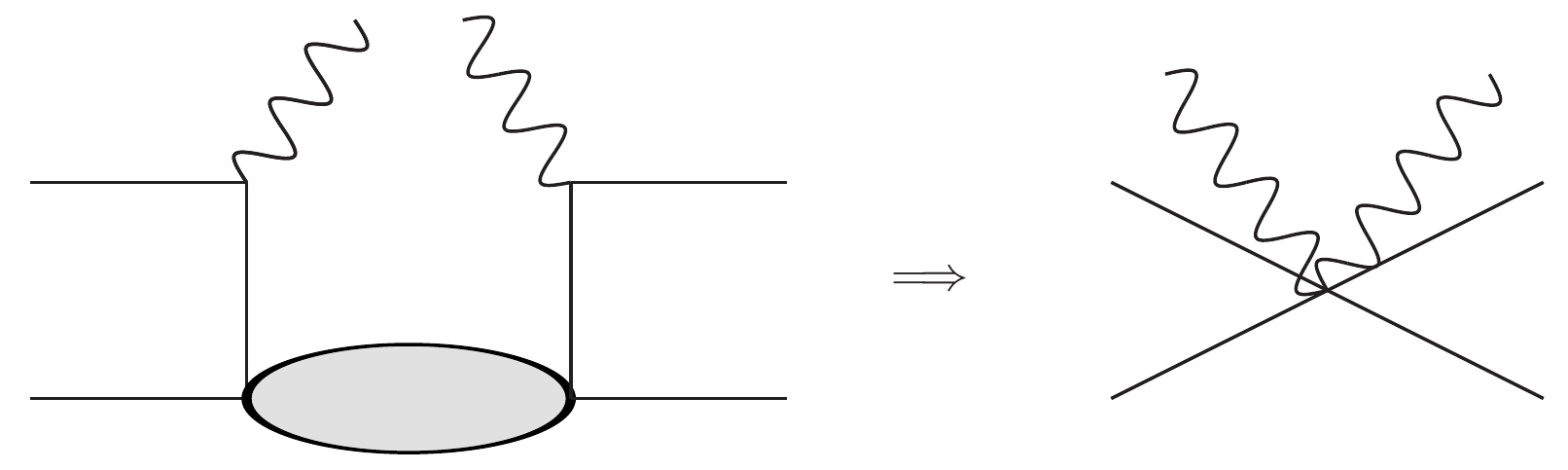}}}
\vskip-0.3cm
\caption[1]{This diagrams illustrates how the hard modes are integrated out (the blob) to generate a six body operator.
The wavy lines are photons. } 
\label{hard}
\vskip-0.5cm
\end{figure}

When integrating out the hard modes whose invariant mass squared is of order $\lambda M_\chi^2$, one generates a set of operators which will be responsible for WIMP annihilation.  
Here we will concentrate on photonic final states, as that is most directly 
relevant for indirect detection. Thus, our observable of interest is the semi-inclusive annihilation rate, $\chi^0 \chi^0 \rightarrow \gamma + X$.  Since the state recoiling against the photon is general, and generically off-shell by an amount $\lambda M_\chi^2$, we can use the operator product expansion (OPE) to work directly with operators whose expectation values give the semi-inclusive matrix element-squared for our process of interest.  This is illustrated in Figure \ref{hard}.  Furthermore, we will consider the annihilation of Majorana
WIMPs in the adjoint, which in the MSSM corresponds to the gauginos (extension to Higgsino DM is straightforward). The minimal operator basis is then
given by
\bea
O_1&=& \left( \bar  \chi \gamma^5 \chi \right) \vacproj \left( \bar \chi \gamma^5 \chi \right) B^{\mu A \perp} B_\mu^{ A \perp}\nn\\
O_2&=& \frac{1}{2}\Big\{\left( \bar \chi \gamma^5 \chi \right) \vacproj \left( \bar \chi_A \gamma^5 \chi_B \right) \nn \\
&&+ \left(\bar \chi_A \gamma^5 \chi_B \right) \vacproj \left( \bar \chi \gamma^5 \chi \right) \Big\} B_\mu^{\perp A} B^{\mu B \perp}\nn\\
O_3&=& \left( \bar \chi_C\gamma^5 \chi_D \right) \vacproj \left( \bar \chi_D \gamma^5 \chi_C \right) B^{\mu A \perp} B_\mu^{ A \perp}\nn\\
O_4&=& \left( \bar \chi_A \gamma^5 \chi_C \right) \vacproj \left( \bar \chi_C \gamma^5 \chi_B \right) B_\mu^{\perp A} B^{\mu B \perp},
\label{ops}
\eea
where we use the vacuum insertion approximation in the WIMP sector, which is valid up to $\mo(v^2)$ corrections.  Henceforth, we drop the explicit vacuum projector.  Implicitly, there is also a projection onto a single-photon state between the $B^{\mu \perp}$ fields i.e.
\beq
 B_\mu^{\perp A} B^{\mu B \perp}\equiv \sum_X  B_\mu^{\perp A}\mid \gamma+X\rangle\langle \gamma+X\mid B^{\mu B \perp}.
\eeq
The $\chi$ operators have implicit projectors onto their large components.
All operators which arise in the  matching can be reduced to one of these four using
the Majorana condition. The spin one operators are irrelevant 
since Fermi statistics would lead to an anti-symmetric SU(2) initial state, and we are interested
in the annihilation of two neutral particles. We have also used the definition
\beq
B^{A \perp}_\mu \equiv f^{ABC}\, W_n^T (D^\perp_\mu)^{BC} \, W_n ,
\eeq
where the $\perp$ symbol implies the component perpendicular to the large light cone momentum $n \cdot p$, without loss of generality $n^\mu=(1,0,0,1)$, and $D^\perp_\mu$ is the covariant derivative
in the collinear sector (for details see \cite{SCET}).
This field interpolates for a collinear gauge boson and is invariant under collinear gauge transformations
due to the Wilson lines on both sides,  $W_n= P \exp \left[ig\int_{-\infty}^0  n \cdot A_n(n\lambda)  d\lambda \right]$,
where $A_n(x)$ is the collinear SU(2) gauge boson field with large momentum $n \cdot p$.
These operators are dressed by identical soft Wilson lines such
that $O_2$ and $O_4$ become
\bea
O_2&=& \frac{1}{2}\Big\{\left( \bar \chi \gamma^5\chi \right) \left( \bar \chi_{A^\prime} \gamma^5 \chi_{B^\prime} \right) + \left( \bar \chi_{A^\prime} \gamma^5 \chi_{B^\prime} \right) \left( \bar \chi \gamma^5\chi \right) \Big\} B^{\tilde A} B^{\tilde B} \nn\\
&&S_{v A^\prime A}^T S_{vB B^\prime}S_{n  \tilde A A}^TS_{n B \tilde B}\nn\\
O_4&=& \left( \bar \chi_{A^\prime} \gamma^5 \chi_C \right) \left( \bar \chi_C \gamma^5 \chi_{B^\prime} \right)  B^{\tilde A} B^{\tilde B} S_{v A^\prime A}^T S_{vB B^\prime}S_{n  \tilde A A}^TS_{n B \tilde B}.\nn\\
\eea
As we can see there are two types of path ordered soft Wilson lines $S_v$ and $S_n$ defined by
\bea
S_{(v,n)}&=&P[ e^{ig \int_{-\infty}^0 (v,n) \cdot A((v,n)\lambda) d\lambda}],
\eea
where here $A$ is the soft gluon.
The operators $O_{1,3}$ receive no soft corrections.   We will be interested in a particular matrix element of these operators.
Taking the common product of soft Wilson lines in $O_{2,4}$ as $O^a_s \equiv S_{v A^\prime A}^T S_{vB B^\prime}S_{n  \tilde A A}^TS_{n B \tilde B}$, the annihilation spectrum may be written as
\bea
\label{cross}
&& \frac{1}{E_\gamma}  \frac{d\sigma}{dE_\gamma} = \frac{1}{4M_\chi^2 v} \langle 0 \mid O^a_s \mid 0 \rangle \nn \\
&\times& \Bigg[ \int d n \cdot p \, \Bigg\{ C_2(M_\chi, n \cdot p) \langle p_1 p_2 \mid \frac{1}{2}\Big\{\bar \chi \gamma^5\chi \, \bar \chi_{A^\prime} \gamma^5 \chi_{B^\prime} \nn \\
&+& \bar \chi_{A^\prime} \gamma^5 \chi_{B^\prime}\bar \chi \gamma^5\chi \Big\}(0) \mid p_1 p_2 \rangle + C_4(M_\chi, n \cdot p) \nn \\
&\times&\langle p_1 p_2 \mid \bar \chi_{A^\prime} \gamma^5 \chi_C \, \bar \chi_C \gamma^5 \chi_{B^\prime} (0) \mid p_1 p_2 \rangle \Bigg\} F^\gamma_{\tilde A \tilde B}\left( \frac{2E_\gamma}{n \cdot p} \right) \Bigg] \nn\\
&&+  \Bigg[ \int d n \cdot p \,\Bigg\{ C_1(M_\chi, n \cdot p) \nn \\
&\times& \langle p_1 p_2 \mid \bar \chi \gamma^5 \chi \, \bar \chi \gamma^5 \chi (0) \mid p_1 p_2 \rangle +C_3(M_\chi, n \cdot p) \nn \\
&\times& \langle p_1 p_2 \mid \bar \chi_C \gamma^5 \chi_D \, \bar \chi_D \gamma^5 \chi_C (0) \mid p_1 p_2 \rangle\Bigg\} F_\gamma\left( \frac{2E_\gamma}{n \cdot p} \right)  \Bigg] , \nn \\
\eea
where the $F^\gamma_{\tilde A \tilde B}$ is a fragmentation function defined by
\bea
F^\gamma_{\tilde A \tilde B}\left( \frac{n\cdot k}{n\cdot p} \right) &=& \int \frac{ dx_-}{2\pi}e^{in \cdot p x_-}\langle 0 \mid B^{\perp \mu}_{\tilde A}(x_-) \mid \gamma(k_n)+X_n \rangle \nn \\
&\times& \langle \gamma(k_n)+X_n \mid  B^\perp _{\mu \tilde B}(0)  \mid 0 \rangle ,
\eea 
and $F_\gamma = F^\gamma_{\tilde A \tilde B} \delta_{\tilde A \tilde B}$.
Note that this is an unusual fragmentation function in that we are measuring states which are not gauge singlets.
Of course, our initial states are not singlets either. 
$C_i$ are the matching coefficients that give the probability 
for the dark matter to annihilate and create a photon with momentum $ n \cdot p$.
$F^\gamma$ is the canonical fragmentation function giving the probability of an initial photon with momentum $p$ to yield
a photon with momentum fraction $n\cdot k /n \cdot p$ after splitting. Since the $C_{1,3}$ contributions in Eq.~(\ref{cross})
are not sensitive to the nonsinglet nature of the initial state, it will only contribute large double logs from mixing with $O_{2,4}$. 

In writing down Eq.~\ref{cross}, we factorized the collinear and soft fields, as the total Hilbert 
space of the system is a tensor product of the soft and collinear sector. The potentials which determine the four quark operator matrix element will in general talk to
the soft sector. However, these interactions will not lead to large double logs.

The large logs are summed by the running of these operators from the scale $M_\chi$ to the scale $M_W$.
This running is noncanonical in that it involves both renormalization group (RG) as well as rapidity renormalization
(RRG) \cite{RRG} group running. The canonical fragmentation function does not have rapidity divergences as there are cancellations between
the real and virtual emissions. However, $F^\gamma_{\tilde A \tilde B}$ will have rapidity divergences due to
the nonsinglet nature of the measured particle. These divergences will only cancel once one accounts
 for the rapidity divergences which arise in the soft part of the operator $O_2$. This provides a nontrivial
 calculational check. Thus in addition to the renormalization scale $\mu$ there is also 
a rapidity scale $\nu$.  Each of the three components of the factorized rate sit at a natural scale
for both $\mu$ and $\nu$. 
The soft and collinear  sectors have no large logs if we choose the $(\mu,\nu)$ scales to be $(M_W,M_W)$ and
 $(M_W,M_\chi)$ respectively. At leading double log accuracy we can resum all of the relevant terms
 by choosing $\mu=M_W$. In this case all the large logs reside in the renormalized parameter $C_{i}(\mu=M_W)$
 and the rapidity running may be neglected.
 
\begin{figure}
\centerline{\scalebox{0.45}{\includegraphics{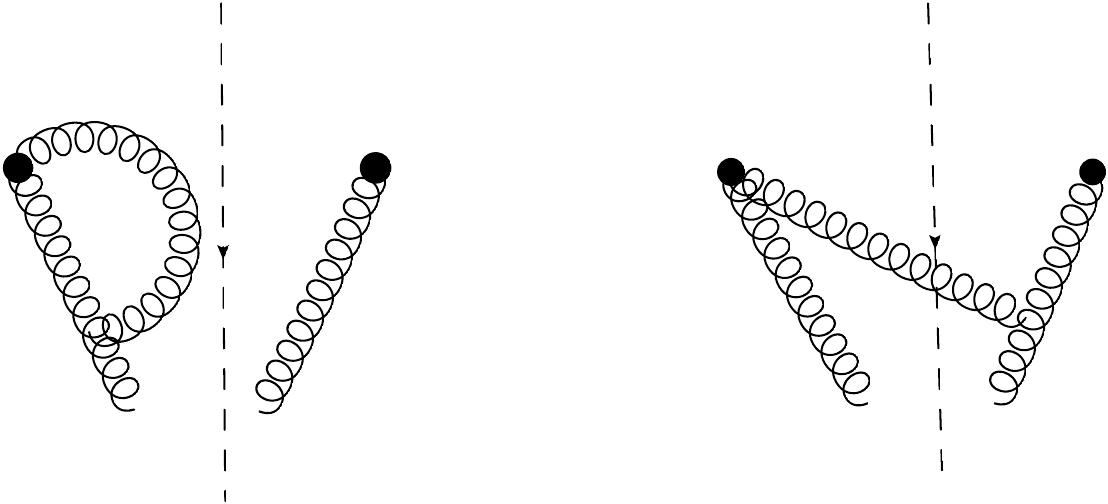}}}
\vskip-0.2cm
\caption[1]{The two diagrams which lead to rapidity divergences in the fragmentation function.
The dashed line represents the cut throughout which final states pass. The solid dot represents
the gauge invariant field strength $B_\mu^\perp$.}
\label{coll} 
\vskip-0.3cm
\end{figure}

\begin{figure}
\centerline{\scalebox{0.5}{\includegraphics{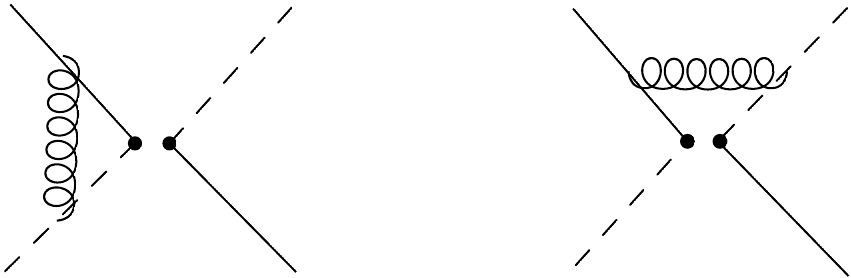}}}
\vskip-0.2cm
\caption[1]{Diagrams which contribute rapidity divergences to the soft factor. The dashed/solid line represents
the time/light-like Wilson line. }
\label{soft} 
\vskip-0.3cm
\end{figure}

To calculate the anomalous dimensions we first introduce an operator basis in the collinear and soft sectors
\bea
O_s^a&=&S_{v A^\prime A}^T S_{vB B^\prime}S_{n  \tilde A A}^TS_{n B \tilde B}~~~~~~~~~O_s^b=\mathbb{1} \, \delta_{\tilde A \tilde B} \delta_{A^\prime B^\prime}
\nn \\
O_c^a &=& B^\perp_{\tilde A}B^\perp_{\tilde B}~~~~~~~~~~~~~~~~~~~~~~~~~O_c^b= B^\perp\cdot B^\perp \delta_{\tilde A \tilde B}.
\eea
These operators each mix within their respective sectors via the diagrams shown in Figs.~\ref{coll} and \ref{soft}.
such that,
\beq \mu \frac{d}{d\mu}\left(\begin{array}{c} O^{c,s}_a \\ O^{c,s}_b \end{array}\right)=\left( \begin{array}{cc} \gamma^{c,s}_{aa} & \gamma^{c,s}_{ab}  \\ 0 & 0 \end{array} \right) \left(\begin{array}{c} O^{c,s}_a \\ O^{c,s}_b \end{array}\right). \eeq
The anomalous dimensions are given by
\bea
\gamma^c_{aa} &=& \frac{3g^2}{4\pi^2} \log(\frac{\nu^2}{4M^2_\chi}), \;\;\; \gamma^s_{aa} = \frac{-3g^2}{4\pi^2} \log(\frac{\nu^2}{\mu^2}), \nn \\
\gamma^c_{ba} &=& \frac{-g^2}{4\pi^2} \log(\frac{\nu^2}{4M^2_\chi}), \;\; \gamma^s_{ba} = \frac{g^2}{4\pi^2} \log(\frac{\nu^2}{\mu^2}). 
\label{ads}
\eea

These results allow us to read off the running of the hard matching coefficients $C_{1,2}$  by imposing that the cross section be RG invariant. This leads to  the set of equations
\bea
\mu \frac{d}{d\mu}C_{2,4}(\mu) &=& - (\gamma^c_{aa}+\gamma^s_{aa}) C_{2,4} \nn \\
\mu \frac{d}{d\mu}C_{1,3}(\mu) &=& - (\gamma^c_{ba}+\gamma^s_{ba}) C_{2,4}. 
\label{wilsonrg}
\eea
Notice that the RHS of Eq.~\ref{wilsonrg} is independent of the rapidity scale as it must be.

Solving the RG equations gives the following result for the resummed cross section in terms of the square of the wave functions at the origin, which are equal to the matrix elements of the four Fermi operators up to corrections
in the relative velocity,
\bea
\frac{1}{E_\gamma}\frac{d\sigma}{dE_\gamma} &=& \frac{C_1(\mu=E_{\gamma})}{4M^2_{\chi} \, v}\delta(E_\gamma-M_\chi) \left[ \frac{2}{3}f_- \mid \! \psi_{00}(0)\!\mid^2 \right. \nn \\
&+& \left. 2 f_+\mid \! \psi_{+-}(0)\!\mid^2 
+ \frac{2}{3}f_-(\psi_{00}\psi_{+-}+{\rm h.c.}) \right]
\label{final}
\eea
where  $f_\pm \equiv 1\pm\exp[- a \log^2(\frac{M_W}{E_{\gamma}})]$, and the factor dividing $C_1$ accounts for the photon polarization sum and the dark matter flux factor.  The $\delta$-function is a consequence of matching for the Wilson coefficient at tree level. The wavefunction-at-the-origin terms are schematically $\psi = \langle p_1 \, p_2 | \bar \chi_A \gamma^5 \chi_B |0 \rangle$, and we insert the color structures of operator basis Eq.~\ref{ops} and rotate into mass eigenstate basis.  They account for the long-distance physics responsible for Sommerfeld Enhancement as well as the ability for the asymptotic neutral state, $\chi^0 \chi^0$ to annihilate through mixing with the charged state, $\chi^+ \chi^-$.  To present a model independent form we have used the fact  that the tree level result for $\chi^0 \chi^0 \rightarrow \gamma + \gamma/Z$ must vanish  in which case there is only one independent matching coefficient.  The tree level result for the
fragmentation function as well as for the soft Wilson line have been used, which is sufficient at leading
double log accuracy.
The anomalous dimensions fix $a=\frac{3\alpha}{\pi}$ for the exponent in the Sudakov factors, $f_{\pm}$.

%
\begin{figure}[ht]
\centerline{\scalebox{0.3}{\includegraphics{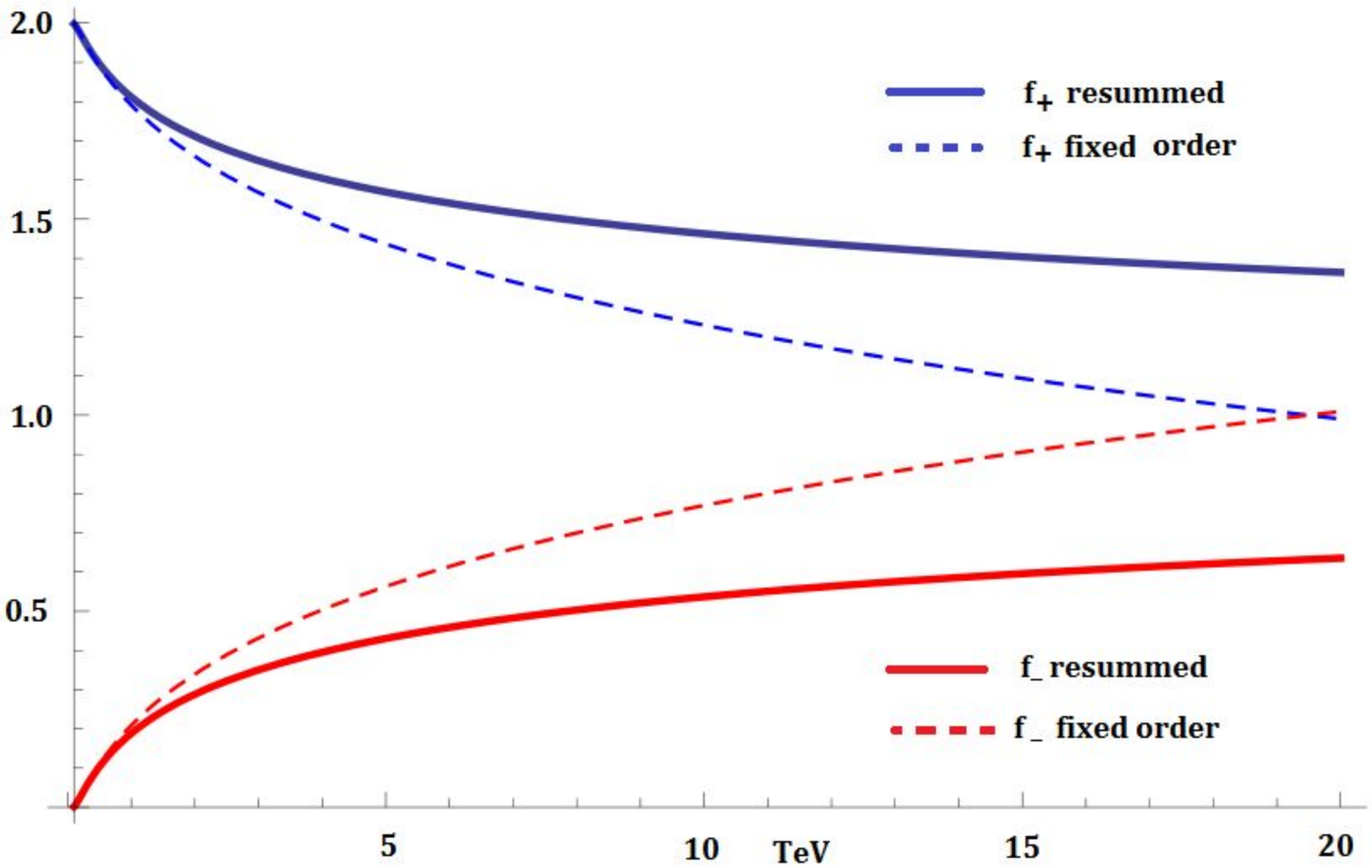}}}
\vskip-0.3cm
\caption{The resummed Sudakov factors $f_+,f_-$ as a function of the neutralino mass $M_{\chi}$. }
\label{fpm} 
\vskip-0.5cm
\end{figure}

The matrix element of the $\chi$ fields, giving the $\psi$ factors, should be evaluated
by solving the Schr\"{o}dinger equation in the presence of the potential created by $W$ exchange (when $\beta\equiv \alpha/(4 \pi) M_\chi^2/M_W^2>1$),  and in general will be model dependent.  However, in the limit that the SU(2) triplet is the only new electroweak state at the TeV scale, then the mass splitting between charged and neutral states is $\approx$ 170 MeV, independent of the specific $M_\chi$ \cite{Cohen:2013ama}.  Furthermore, the Sommerfeld factors $\psi_{00,+-}$ are comparable \cite{tracysomm}, as is unsurprising for a multiplet split at the $10^{-4}$ level.  We can therefore see that the impact of radiative corrections on the annihilation rate $\chi^0 \chi^0 \rightarrow \gamma + X$ will be at the few percent level.  In the limit that only the $f_+$ term contributes, which is equivalent to all annihilation coming from the charged state, we see a modest decrease with increasing $M_\chi$.  For the thermal relic value of 3 TeV, this is $\sim$10\%.  However, the $f_-$ contributions increasingly turn on over the same range (Fig.~\ref{fpm}) and at the same scale.  While we leave the quantification of this effect for future work \cite{future}, we see that the total effect of leading log corrections will be very modest and far from the potential factor-of-a-few suppression claimed in \cite{Cohen:2013ama}.  
 
In this letter we have presented a factorization theorem for the annihilation of Majorana
WIMPs into photons.
This theorem allows for the calculation of  radiative corrections with relative ease, as
the Coulomb physics is disentangled from the final state up to power corrections
in $\lambda=M_W/E_\gamma$. This is to say that the corrections to the factorization
result are of order $\lambda$, which is clearly sufficient at least until the
time a discovery is made.  The final result (\ref{final}) can be utilized to
generate the cross section once a model is chosen within which to evaluate 
the matrix elements of the four fermion operators. The formalism introduced here
can be generalized to the case of WIMPs transforming in the fundamental of
SU(2) with relative ease. We have chosen the Majorana (gaugino-like) case
here since it has a reduced operator basis and has been of recent phenomenological interest.  Beyond the implications 
for dark matter, our formalism allows one to resum the kinematic double logarithms that arise in inclusive observables 
from the radiation of gauge bosons from non-singlet external states in a broken, nonabelian gauge theory.

\vspace{0.1in}

\section*{Acknowledgements}
The authors would like to thank Ambar Jain and Mannie Chiu for early collaboration on this subject.
We also  thank Tracy Slatyer for discussions. The authors are supported by DOE grants DOE DE-FG02-04ER41338 and FG02-06ER41449.
M.B. acknowledges the Aspen Center for Physics where a portion of this work was completed.

\vspace{-0.2in}

\section{Note added}
While this work was in final preparation, we became aware of forthcoming papers on a similar topic: {\it Soft-collinear effective theory for WIMP annihilation} by Martin Bauer, Timothy Cohen, Richard J.~Hill, and Mikhail P.~Solon; and {\it Heavy dark matter annihilation from effective field theory} by Grigory Ovanesyan, Tracy R.~Slatyer, and Iain W.~Stewart.

\end{document}